\documentclass[conference]{IEEEtran}
\IEEEoverridecommandlockouts
\usepackage{cite}
\usepackage{amsmath,amssymb,amsfonts}
\usepackage{algorithmic}
\usepackage{graphicx}
\usepackage{textcomp}
\usepackage{xcolor}
\usepackage{tabularx}
\usepackage{booktabs}
\usepackage{url}

\def\BibTeX{{\rm B\kern-.05em{\sc i\kern-.025em b}\kern-.08em
    T\kern-.1667em\lower.7ex\hbox{E}\kern-.125emX}}
\begin{document}

\title{DiRAC - Distributed Robot Awareness and Consensus\\

}

\author{\IEEEauthorblockN{Kashish Mittal}
\IEEEauthorblockA{\textit{Dept. of CSE} \\
\textit{PES University}\\
kashishmittal6002@gmail.com}
\and
\IEEEauthorblockN{Lakshasri S}
\IEEEauthorblockA{\textit{Dept. of CSE} \\
\textit{PES University}\\
lakshasrishivakumar@gmail.com}
\and
\IEEEauthorblockN{Manjari Kulkarni}
\IEEEauthorblockA{\textit{Dept. of ECE} \\
\textit{PES University}\\   
tattvamanjari0145@gmail.com}
\and
\IEEEauthorblockN{Uday Gopan}
\IEEEauthorblockA{\textit{Dept. of CSE} \\
\textit{PES University}\\
udaygopan04@gmail.com}
\and

\IEEEauthorblockN{Sriram Radhakrishna}
\IEEEauthorblockA{\textit{Dept. of CSE} \\
\textit{PES University}\\
sriram.radhakrishna42@gmail.com}
\and
\IEEEauthorblockN{Aditya Naskar}
\IEEEauthorblockA{\textit{Dept. of CSE} \\
\textit{PES University}\\
adityanaskar1033@gmail.com}
\and

\IEEEauthorblockN{Rameshwar DL}
\IEEEauthorblockA{\textit{Dept. of CSE} \\
\textit{PES University}\\
rameshwar77411@gmail.com}
}

\maketitle

\begin{abstract}
DiRAC is a scalable, distributed framework designed to enable efficient task assignment and path planning in very large robotic swarms. It introduces a novel zone-partitioned architecture with dynamically elected leaders and a tick-synchronized consensus protocol that yields strong consistency and deterministic outcomes. For path planning, DiRAC uses a novel algorithm, a force-based decentralized planner for real-time collision resolution. Validated within ROS 2 middleware through preliminary simulation, DiRAC demonstrates architectural scalability and modular efficiency in simulated warehouse environments, laying the groundwork for real-world deployment in large-scale industrial and logistics domains.
 
\end{abstract}

\begin{IEEEkeywords}
Swarm Robotics, Multi-Agent Systems, Distributed Consensus, Task Assignment, Path Planning, Distributed Algorithms, Robot Coordination, Scalable Systems, Leader Election, Fault Tolerance, Cooperative Control, Decentralized Control, ROS 2 Middleware.
\end{IEEEkeywords}

\section{Introduction}

Large-scale robot swarms are increasingly central to high-throughput autonomy in domains such as warehousing and logistics, where many agents must continuously coordinate to meet real-time performance and safety constraints at scale~\cite{b1}. As these systems grow beyond small teams, centralized control assumptions break down: centralized supervisors become bottlenecks, single points of failure emerge, and maintaining a consistent shared world state across many agents becomes both harder and more essential~\cite{b5}. To address this, agents must be capable of autonomous decision making, allocating tasks among themselves, planning collision-free motion, and reaching state agreement in real time through decentralized self-organization alone, while still producing deterministic, reproducible decisions across a large, homogeneous population~\cite{b6}.

Despite progress, core gaps persist in today’s solutions. Centralized Fleet Management Systems can be optimized globally, but at scale, they create computational and network choke points and concentrate failure risk, limiting robustness in dynamic environments~\cite{b10}.

A particularly challenging aspect of swarm coordination is motion planning, which requires algorithms to generate collision-free paths for multiple agents while balancing computational efficiency and global coherence. Optimal approaches such as Conflict-Based Search (CBS) provide strong guarantees but suffer from exponential degradation as agent density and number increase~\cite{b2}. In contrast, reactive and heuristic methods, such as potential field methods and iterative algorithms like Priority Inheritance with Backtracking (PIBT), which prioritize scalability and responsiveness over complete global optimality, often fall short in providing the precise coordination and state consistency required for dense, real-time multi-agent environments~\cite{b3,b8,b9}. Critically, few frameworks unify job assignment, distributed consensus, and decentralized path planning into a single architecture that ensures a consistent shared state across subsystems in real time.

This paper introduces DiRAC, a framework centered on a tick synchronized consensus mechanism inspired by RAFT with acknowledgments that emphasizes consistency and partition tolerance, ensuring that all agents plan against a shared, agreed state and produce deterministic outcomes given identical input even amid disconnections and recoveries~\cite{b4}. To enable scalable coordination without centralized bottlenecks, DiRAC employs a zone-partitioned architecture with dynamically elected zone leaders for local task allocation and a supervising superleader for global routing and load balancing. This localizes communication, reduces contention, preserves global coherence, and eliminates single points of failure~\cite{b1,b5}. Complementing this consensus core, a novel cooperative, force-based multi-agent path finding algorithm handles vertex, edge, static, and deadlock conflicts at scalable complexity, tightly integrated with consensus and zonal job allocation to deliver end-to-end swarm control capable of managing thousands of agents~\cite{b8}. While this paper introduces DiRAC as a unified framework, subsequent publications will expand on its key modules ie. task allocation, consensus synchronization, and decentralized path planning and moving toward experimental validation.

\section{Related Work}

The DiRAC framework synthesizes principles from four foundational research domains to address the challenges of large-scale autonomous logistics: decentralized control architectures, distributed consensus, Multi-Agent Path Finding (MAPF), and dynamic fleet management. While substantial research exists within each domain, their integration into a cohesive, real-time system for robotic swarms remains an open problem. DiRAC is designed to bridge this gap.

\subsection{Decentralized Control Architectures}

This is critical in enabling the scalability and resilience required by large agent populations. Traditional centralized systems suffer from inherent limitations, including computational bottlenecks and a single point of failure, rendering them fragile and impractical for dynamic, large-scale deployments. In contrast, decentralized topologies distribute computational load and decision-making authority, yielding systems that are robust to individual agent failures and can scale gracefully. Hierarchical decentralization, which introduces local leadership structures, further optimizes performance by confining communication and coordination to relevant subgroups, a core tenet of DiRAC's zoned operational model~\cite{b5}\cite{b6}\cite{b1}.

\subsection{Distributed Consensus}

This provides a mechanism for maintaining a coherent shared state across a decentralized system. Without a reliable method for agents to agree upon a common operational picture, the swarm would devolve into uncoordinated, chaotic behavior. Protocols such as RAFT offer a formal basis for ensuring strong consistency, typically by prioritizing it over system availability~\cite{b4}. For robotic swarms, consensus is not merely a data-level abstraction; it is fundamental to safe, coordinated action, ensuring all agents operate on a validated, unified understanding of their peers' states and intentions~\cite{b7}.

\subsection{Multi-Agent Path Finding (MAPF)}

This addresses the fundamental problem of navigating multiple agents through a shared space without collision. While classical MAPF solvers can derive provably optimal paths, their computational complexity scales exponentially with the number of agents, making them intractable for real-time applications. Consequently, the field has shifted towards scalable, sub-optimal algorithms that prioritize computational tractability and rapid, reactive path planning. Methods employing heuristics (e.g., Priority Inheritance with Backtracking) or physics-based force models achieve the real-time performance necessary for dynamic environments where immediate, conflict-free movement is more critical than theoretical path optimality~\cite{b8}\cite{b9}.

\subsection{Dynamic Fleet Management}

This constitutes the strategic layer of swarm coordination, governing task allocation and resource distribution to maximize system throughput. Auction and cost-based mechanisms are widely used for decentralized task assignment, allowing agents to bid for jobs based on localized metrics such as proximity or available capacity. This must be coupled with continuous load balancing to prevent agent oversaturation in high-demand zones and starvation in others. By dynamically redistributing agents in response to fluctuating workloads, the system ensures efficient resource utilization and minimizes total latency of job completion, which is the primary objective of a logistics swarm~\cite{b10}\cite{b13}.

\section{Methodology}
DiRAC is a decentralized framework enabling real-time job assignment and path planning via collective swarm awareness and consensus. Within this system, three types of robotic entities operate on the map:
 Agents: Robots with standard functionalities responsible for executing tasks and movement.
 Leaders: Agents with extended roles, including acting as the point of contact for job assignment within their zone and leading consensus processes locally.
 Super-Leaders: Specialized agents responsible for global coordination tasks such as orchestrating leader elections across zones and issuing migration mandates for load balancing.
 
\subsection{Partitioned Map}
The environment in which the agents exist is described with static obstacles and boundaries using a discrete 2D map. This map contains 2 types of entities, namely static obstacles and agents. Every agent stores a copy of the map with the static obstacles mapped to plan their initial paths. This map is partitioned into zones to allow for localized collision avoidance and job assignment. An overlap parameter is introduced to avoid inter-zone collisions. Agents are subscribed to update topics associated with the zone in which they are currently present, as well as any neighboring zones if they are in the overlapping region. Each agent independently computes its association with a zone, and subsequent computations for operation are performed using information from other agents and obstacles in the same zone. This zoning allows for distributed computation of paths and job assumption costs, as conventional graph path planning algorithms scale poorly with global agent count~\cite{b14}. A leader is elected from among the agents in the zone by the super-leader. This election is based on the proximity of agents to the center of the zone. Agents compute their distance to the centroid of the zone and send this information to the super-leader, who assigns the role of leader to the agent closest to the center.
\[
G = \left( \frac{x_0 + x_1}{2},\; \frac{y_0 + y_1}{2} \right)
\]

\[
\text{Zone Leader} = \min_{\text{Agents}(x, y)} \sqrt{(G_x - x)^2 + (G_y - y)^2}
\]

\subsection{Job Assignment}
Jobs are introduced to the swarm by an independent controller. Independence of the controller means that it does not need to know the system’s state; it may introduce a new job by assigning a location on the map as a job and giving it a priority. The job is then taken up by an agent in the same zone as the job that has the minimum cost to complete it. This ``taking up the job'' is done via the leader of the zone where the job spawns, which confirms job assumption and removes the job from the pool. This prevents conflicting job assignments, as the leader is the source of truth for job assignments. Job assignment is based on the cost for each agent to complete the job and is done similarly to leader election. The leader assigns jobs based on computed costs from agents in the zone. Cost is a function of the distance to the job (Manhattan or the shortest distance based on the map). In real-world deployment, cost may include factors such as remaining battery capacity and the number of turns required to reach the job~\cite{b13}.

\subsection{Load Balancing}
Jobs may be spawned anywhere on the map by the controller, without symmetry across zones. To handle this, a load-balancing mechanism provides the necessary number of agents to service jobs in each zone. This is carried out using a daisy chain mechanism, where a zone pulls an agent from one zone and gives an agent to another zone in the direction of the need, reducing the travel distance for any one agent. The super-leader monitors populations and job counts in each zone, issuing migration mandates to move agents from relatively idle zones to zones with high demand. This load-balancing mechanism optimizes system efficiency by minimizing job waiting times and reducing the time required for agents to complete assigned jobs, thereby decreasing the overall makespan of job batches. It also prevents job starvation. Deadlocks, which may arise in localized load balancing, are avoided by having a single agent issue migration mandates to zones~\cite{b13}.

\subsection{State Consensus}
\begin{figure}[htpb]
  \centering
  \includegraphics[width=0.96\linewidth]{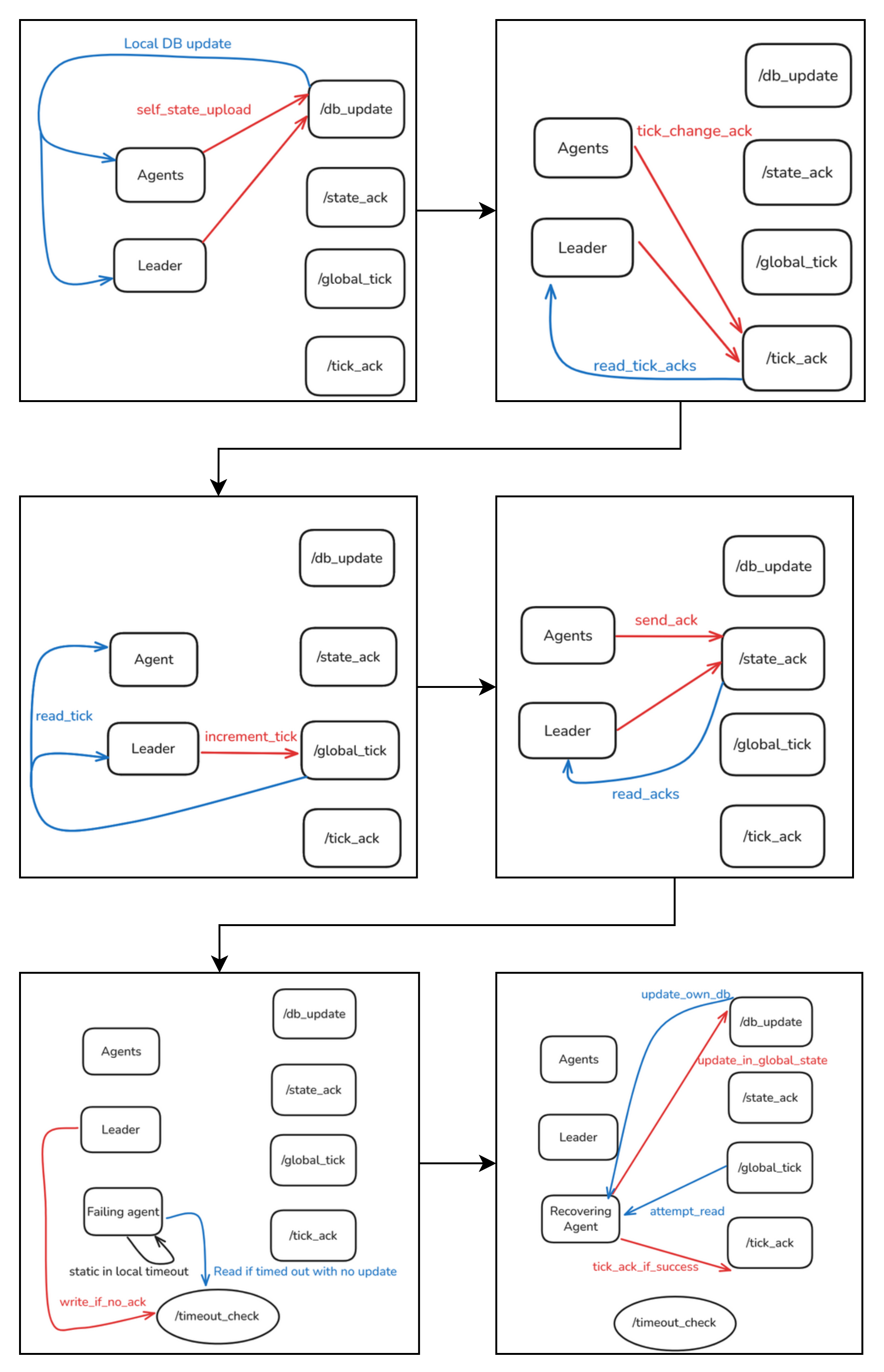}
  \caption{Database State Consensus Mechanism}
  \label{fig:image1}
\end{figure}

State consensus is necessary for all operations in the swarm. A decentralized swarm is a distributed system requiring each node to know the state of every other node in its locality. DiRAC’s state consensus mechanism is a RAFT-inspired strongly consistent algorithm using a logical clock to ensure agents share the same logical tick. Acknowledgements and tick propagation enable strong consistency. Agents that are no longer alive can rejoin the swarm using recovery mechanisms that bring them to the relevant locality’s tick. The state of being ``not alive'' is determined by the zone leader, who is responsible for advancing the tick and may mark a node as ``dead'' if it fails to respond within a predetermined timeout. This recovery mechanism enables partition tolerance. Availability is sacrificed in favor of consistency to avoid frequent collisions from outdated state knowledge. Therefore, a CP system~\cite{b4} is implemented. Conflicts from receiving multiple ticks in overlapping zones do not need to be resolved, since ticks are used to calculate collisions and avoidance among agents of the same zone. In the event of a collision in overlapping regions, the tick of the higher-priority agent is used. Fig.\ref{fig:image1} demonstrates the working of the per tick state-based consensus mechanism. Agent's own state updates are published to the zonal \texttt{db\_update} topic. Agents acknowledge reception of states on the \texttt{/state\_ack}. The \texttt{global\_tick} topic receives the updated tick value from the leader. Once all the agents have acknowledged the tick change, they publish their acknowledgment in the \texttt{tick\_ack} topic, and the leader reads to look for missing acks. Timeout handling and re-synchronization followed by the Tick synchronization process if the agent recovers

\subsection{Path Planning}
DiRAC employs a novel adaptive, force-based cooperative solver for Multi-Agent Path Finding (MAPF). This sub-optimal algorithm prioritizes speed over optimality. It detects multiple collision types, ie, vertex, edge, static, and deadlock, based on agents’ intents and current positions, applying tailored resolutions. Conflicts are resolved through force vectors computed from job locations, agent positions, and intents, scaled by agent priority. Deadlock resolution uses force ramping.  

\subsubsection*{Mathematical Model for Agent Interaction}

The force $\vec{F}$ on an agent $q_i$ is defined by a piecewise function based on the interaction state.
\begin{equation}
\vec{F}(q_i) =
\begin{cases}
    f \cdot P(q_i) \cdot \vec{I}(q_j) + f \cdot P(q_j) \cdot \vec{R}(q_i), & \text{if collision} \\[2ex]
    f \cdot P(q_i)^s \cdot \vec{I}(q_j) + f \cdot P(q_j) \cdot \vec{R}(q_i), & \text{if deadlock} \\[2ex]
    f \cdot (\vec{I}(q_i) - \vec{C}(q_i)), & \text{if no collision}
\end{cases}
\label{eq:force}
\end{equation}

The state is determined by the following conditions, where $\vec{I}(q)$ is the next intended node and $\vec{C}(q)$ is the current node:
\begin{align}
    \vec{I}(q_i) = \vec{I}(q_j) \text{  } &\Rightarrow \text{  Vertex Collision} \label{eq:vertex}\\[1ex]
    \vec{I}(q_i) = \vec{C}(q_j) \land \vec{I}(q_j) = \vec{C}(q_i) \text{  } &\Rightarrow \text{  Edge Collision} \label{eq:edge}\\[1ex]
    \vec{I}(q_i) = \vec{C}(q_j) \land \vec{I}(q_j) = \vec{C}(q_j) \text{  } &\Rightarrow \text{  Static Collision} \label{eq:static}\\[1ex]
    \vec{I}(q_i) = \vec{C}(q_i) \text{  } &\Rightarrow \text{  Wait State} \label{eq:wait}\\[1ex]
    \text{Otherwise  } &\Rightarrow \text{  No collision} \label{eq:nocollision}
\end{align}

\begin{figure}[htbp]
\centering
\begin{minipage}{0.24\textwidth}
\centering
\includegraphics[width=\textwidth]{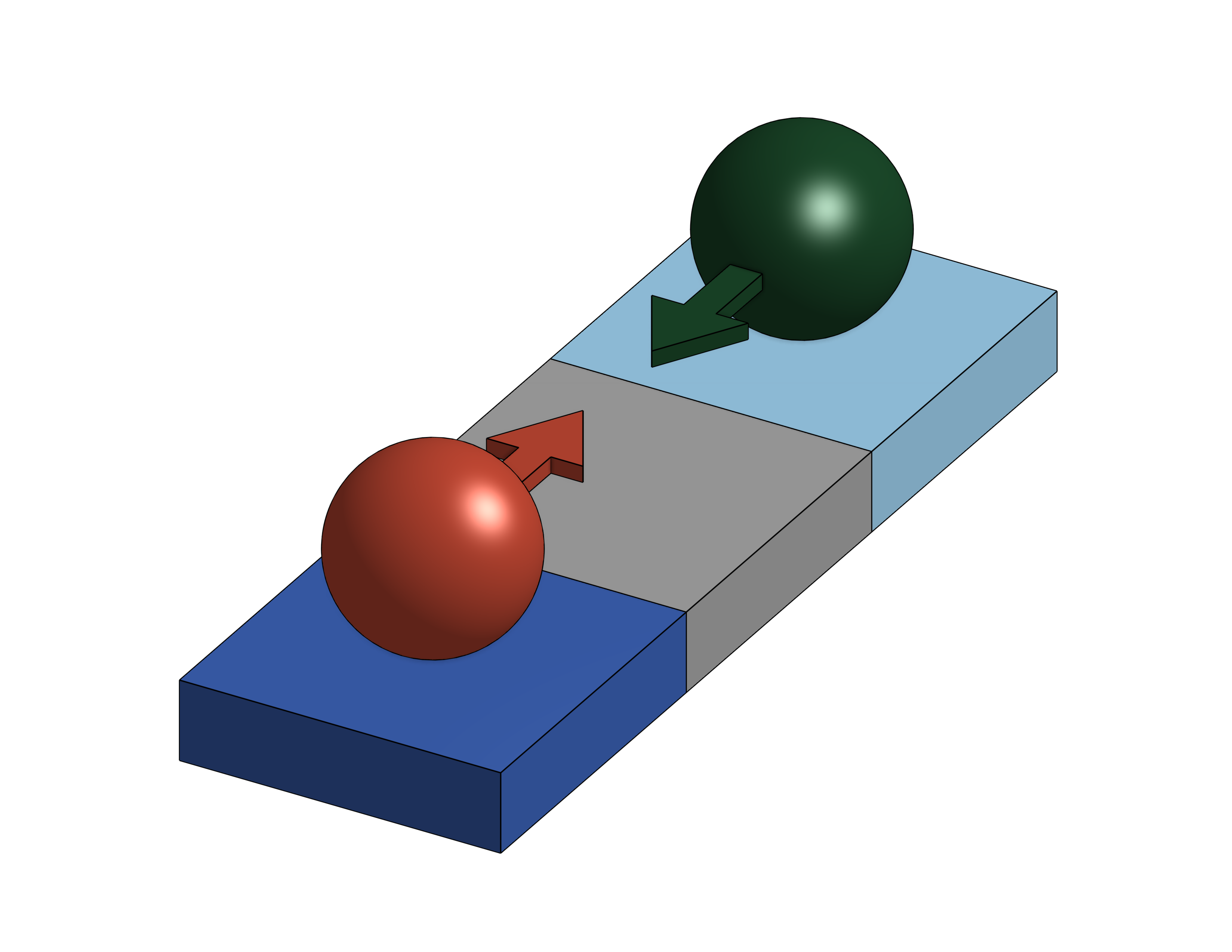}\\
{\small (a) Vertex Collision}
\label{fig:vertex}
\end{minipage}
\hfill
\begin{minipage}{0.24\textwidth}
\centering
\includegraphics[width=\textwidth]{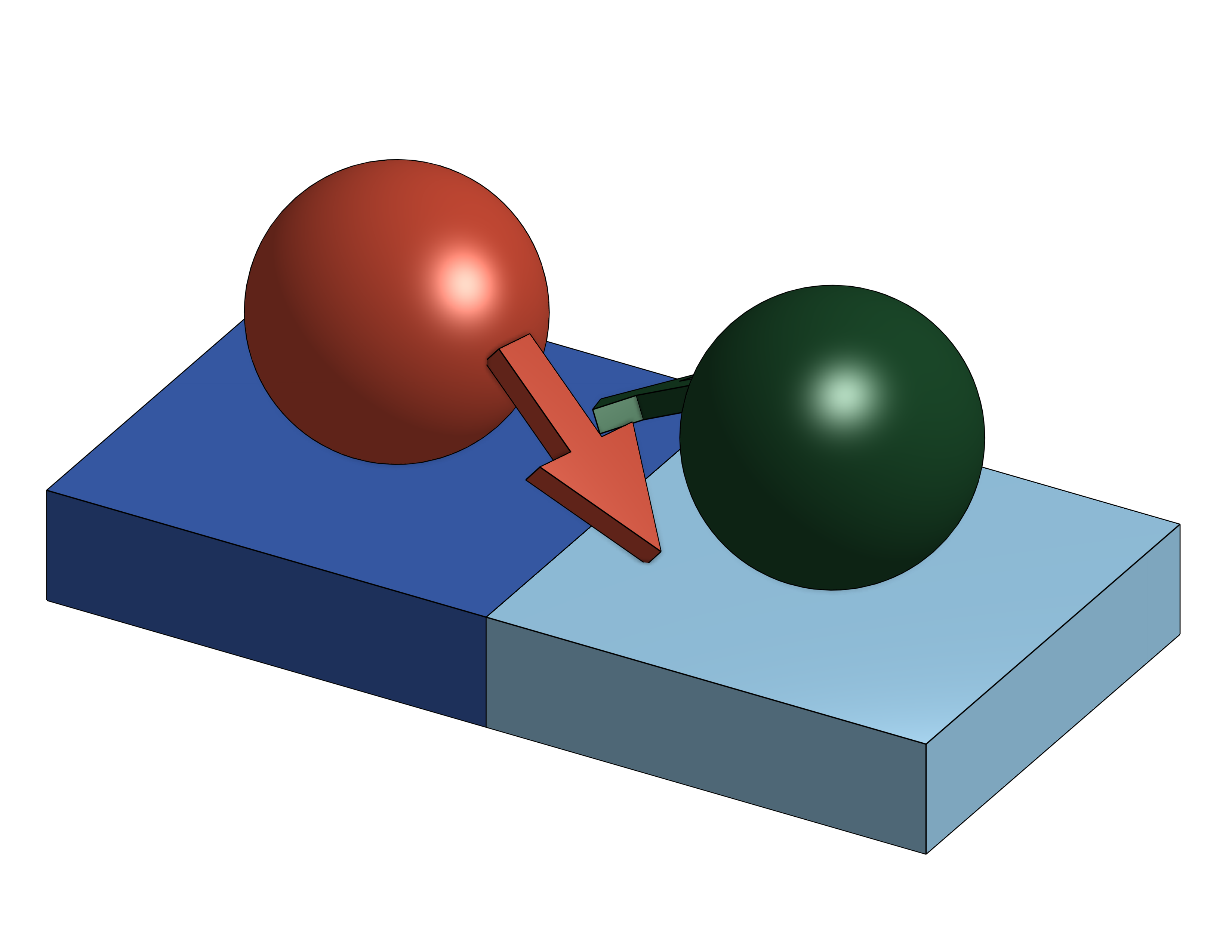}\\
{\small (b) Edge Collision}
\label{fig:edge}
\end{minipage}

\vspace{0.3em}

\begin{minipage}{0.24\textwidth}
\centering
\includegraphics[width=\textwidth]{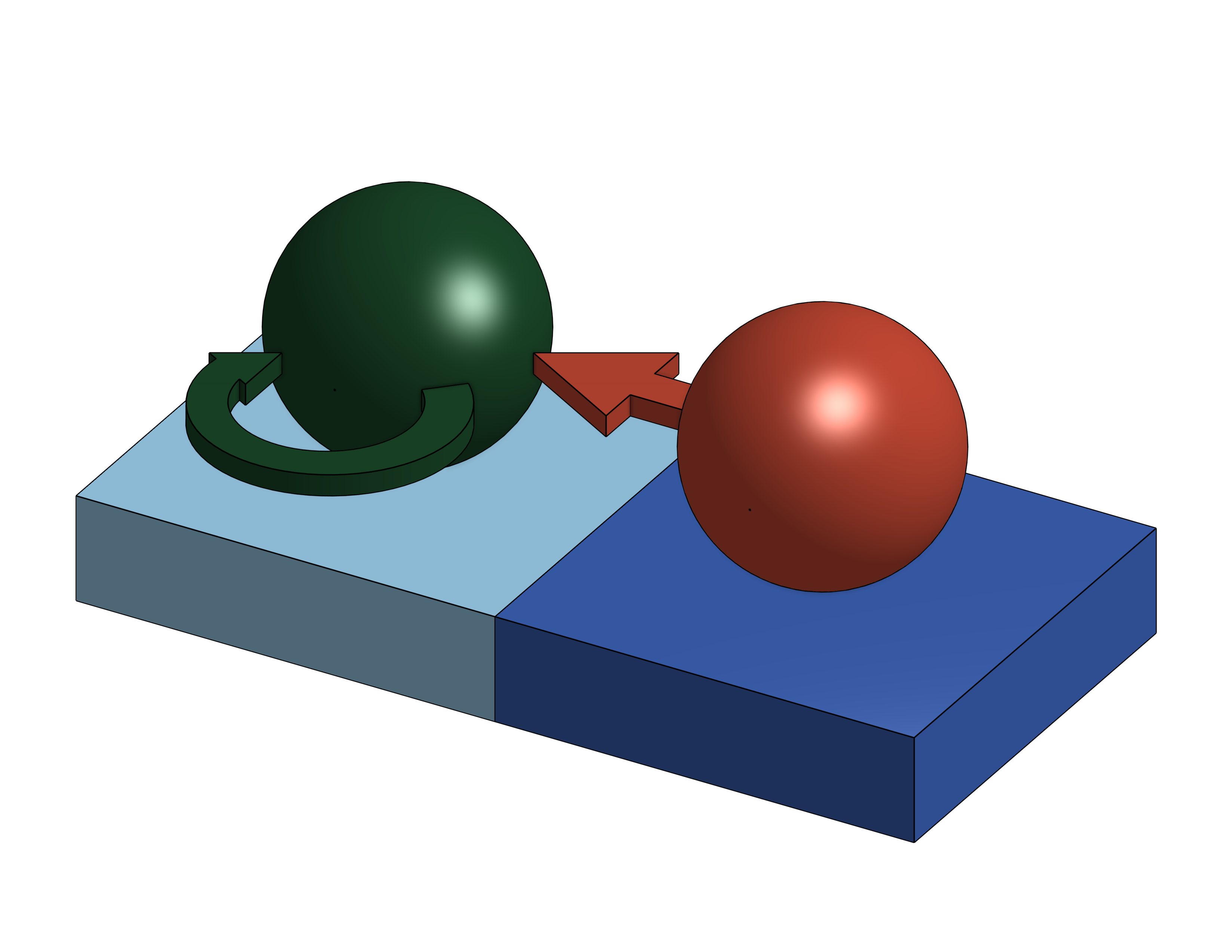}\\
{\small (c) Static Collision}
\label{fig:static}
\end{minipage}
\hfill
\begin{minipage}{0.24\textwidth}
\centering
\includegraphics[width=\textwidth]{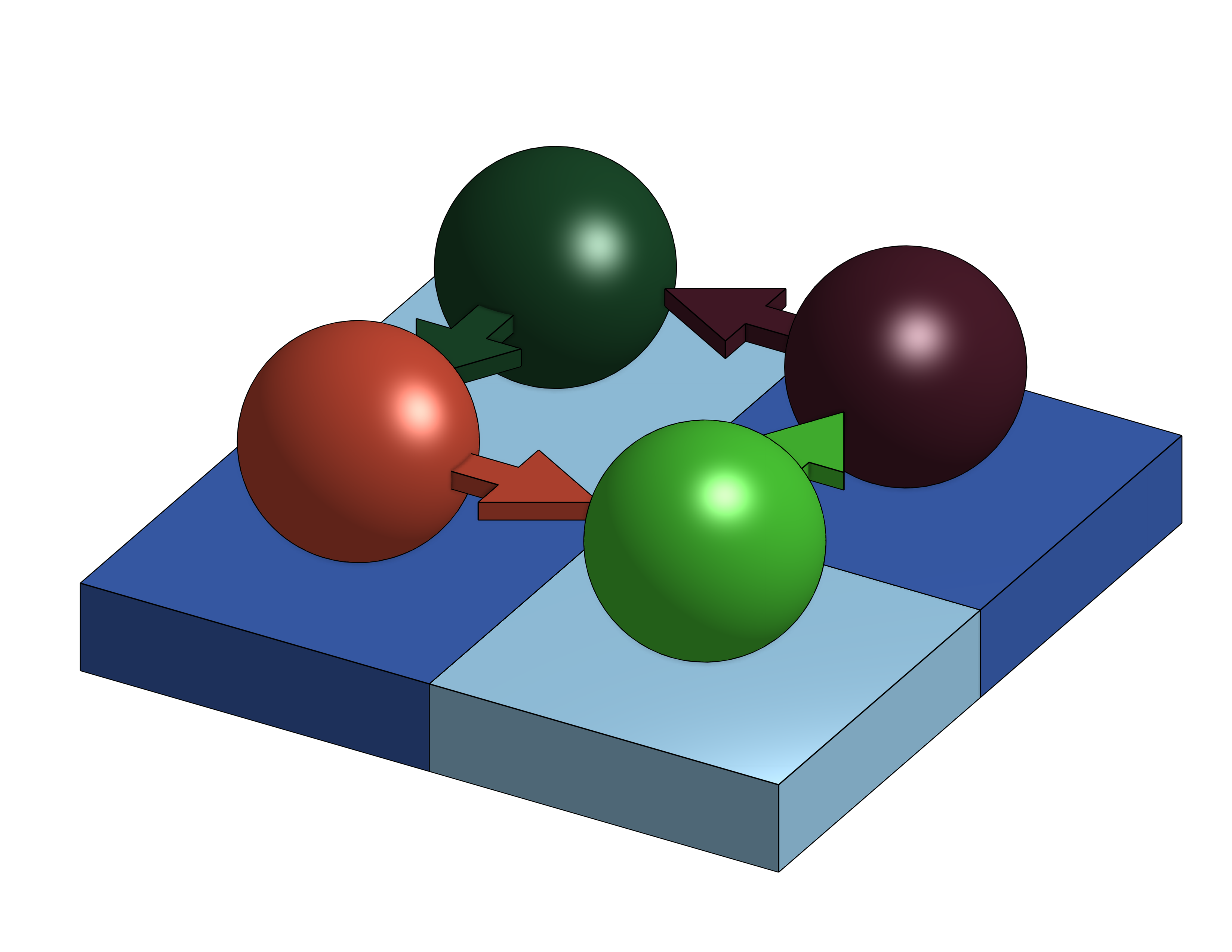}\\
{\small (d) Deadlock}
\label{fig:deadlock}
\end{minipage}

\caption{Illustrations of collision types detected and resolved in the force-based path-planning algorithm.}
\label{fig:collisions}
\end{figure}

\subsubsection*{Variable Definitions}

\begin{align*}
    q_i,\, q_j & : \text{Agent} \\
    \vec{F}(q_i) & : \text{Resultant force vector on } q_i \\
    f & : \text{Base force magnitude (scalar)} \\
    P(q_i) & : \text{Force multiplier of } q_i \text{ (priority)} \\
    s & : \text{Stuck counter, for exponential force scaling} \\
    \vec{I}(q_i) & : \text{Intent (next desired position) of } q_i \\
    \vec{C}(q_i) & : \text{Current position of } q_i \\
    \vec{R}(q_i) & : \text{Random safe vector $q_i$ can take to avoid collision}
\end{align*}

The underlying path planner uses A* with Manhattan distance as the heuristic. The algorithm is designed for high scalability and decentralization, especially when the map is partitioned into zones. Its computational complexity is \(O(G \log G)\), where \(G\) is the number of agents in the zone, with a tight bound of \(\Theta(n)\), where \(n\) is the number of agents that can be involved in a conflict. Communication complexity is \(O(n)\)~\cite{b14}.

\begin{figure}[htbp]
  \centering
  \includegraphics[width=0.48\textwidth]{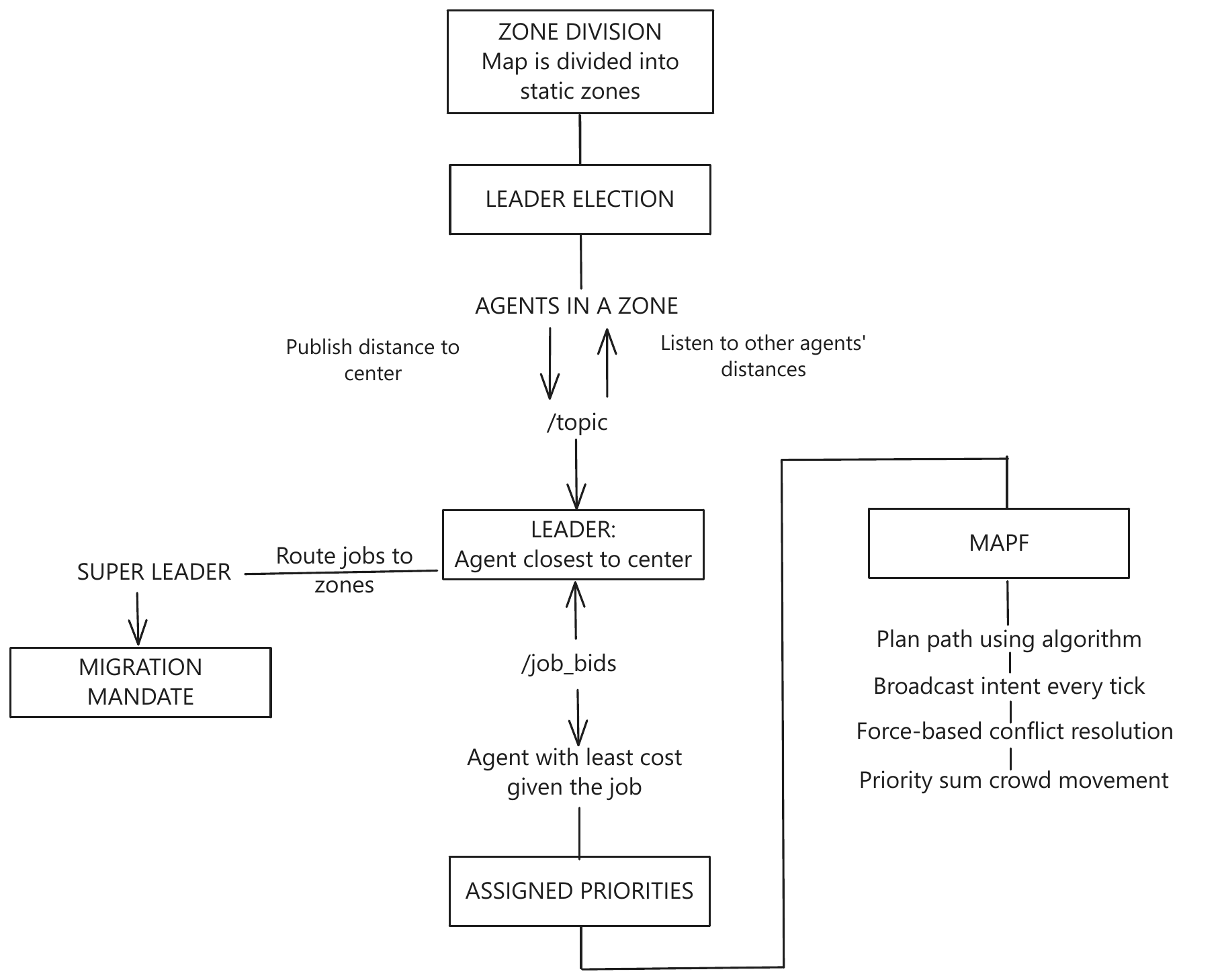}
  \caption{Architecture Diagram}
  \label{fig:my_svg}
\end{figure}

\section{Experimental Setup}

\subsection{Simulation Framework}
The proposed DiRAC framework was implemented and evaluated in a simulated environment using ROS 2 Humble as the middleware~\cite{b11}. Visualization of robot states, zone boundaries, and job assignments was performed through RViz~\cite{b12}. Communication among the agents was realized via the ROS 2 publisher-subscriber model, with custom message definitions for state propagation, leader election, and job assignment.

The system architecture was modularized into independent nodes representing agents, leaders, and the super-leader. Each agent node encapsulated handlers for navigation and task execution. Zone leaders aggregated local information and coordinated job distribution, while the super-leader managed cross-zone coordination, such as leader elections and agent migration for load balancing~\cite{b13}.

The simulation successfully validated leader elections, tick synchronization, and stable inter-zone communication, confirming the viability of the proposed architecture prior to full-scale performance benchmarking

\subsection{Environment}
Experiments were conducted on a \(30 \times 30\) discrete two-dimensional map populated with static rectangular obstacles. The map was partitioned into nine zones with boundary regions to evaluate inter-zone coordination. Initial positions of agents were specified via ROS 2 launch files and remained fixed during initialization. Jobs were introduced by an independent controller node at randomized map coordinates with assigned priorities.

Agent populations of 5, 10, 20, and 30 were tested to study scalability. Job counts varied from 10 to 50 per experiment, with non-uniform distribution across zones to evaluate the load balancing mechanism~\cite{b13}.

\subsection{Framework Comparison}
\label{subsec:framework_comparison}

The framework comparison evaluates architectural strengths across four key attributes relevant to decentralized swarm robotics: Exchange or Communication assesses the support for efficient, decentralized message exchange and state propagation among agents, essential for consensus without central bottlenecks; Concurrency measures the ability to execute multiple tasks or computations simultaneously across agents, enabling scalability in dynamic environments; Coordination evaluates the framework's capability for structured multi-agent collaboration and task distribution, critical for load balancing and team-based operations; and Integration gauges compatibility with existing tools and middleware (e.g., ROS 2 for real-time deployment), facilitating practical implementation and extensibility.

Table~\ref{tab:framework_comparison} qualitatively compares DiRAC's architecture against key recent swarm robotics frameworks focused on decentralized coordination, pathfinding, and task allocation. These architectures emphasize distributed control similar to DiRAC's zone-based consensus and force-based MAPF, but often lack integrated ticking for determinism or full ROS 2 scalability. DiRAC advances by enabling robust, real-time performance across 1000s of agents without simulation dependencies.

\begin{table}[t]
\centering
\small  
\caption{Framework comparison for decentralized swarm architectures.}
\label{tab:framework_comparison}
\begin{tabularx}{0.47\textwidth}{l >{\centering\arraybackslash}X >{\centering\arraybackslash}X >{\centering\arraybackslash}X >{\centering\arraybackslash}X}
\toprule
Framework & Exchange & Conc. & Coord. & Int. \\
\midrule
ALICA~\cite{b18} & $\times$ & $\checkmark$ & $\checkmark$ & $\checkmark$ \\
ROS2swarm~\cite{b15} & $\checkmark$ & $\sim$ & $\sim$ & $\checkmark$ \\
LRP~\cite{b16} & $\checkmark$ & $\times$ & $\checkmark$ & $\times$ \\
Starling (Stanford)~\cite{b17} & $\sim$ & $\checkmark$ & $\checkmark$ & $\sim$ \\
\midrule
DiRAC (Proposed) & $\checkmark$ & $\checkmark$ & $\checkmark$ & $\checkmark$ \\
\bottomrule
\addlinespace
\multicolumn{5}{l}{$\checkmark$: Full support; $\sim$: Partial support; $\times$: Absent.}
\end{tabularx}

\vspace{2pt}
\begin{flushleft}
\footnotesize
Conc.: Concurrency \hspace{1em} Coord.: Coordination \hspace{1em} Int.: Integration
\end{flushleft}
\end{table}

\section{Conclusion and Future Work}

This paper presents DiRAC, a scalable and modular framework for decentralized swarm robotics. Through extensive simulation studies, DiRAC demonstrates robust task allocation, effective leader election, and adaptive multi-agent navigation, even under dynamic conditions and agent dropouts. The system maintains stable and efficient coordination across diverse agent populations and workloads, showcasing its promise for large-scale logistics, automation, and cooperative multi-robot applications. Notably, the architecture delivers reliable consensus and synchronized control with low-latency communication, laying the groundwork for practical deployment in real-world environments.

The architecture presented here establishes DiRAC’s theoretical and software foundations. While large-scale experiments are deferred to future work, the current implementation validates system integrity and modular scalability. Upcoming work will detail the individual subsystems and extend the framework toward real-world deployment \cite{b19}.

Key next steps also involve enhancing scalability through adaptive, density-aware zoning strategies and optimizing the underlying path-planning mechanisms to support even larger, more heterogeneous swarms. These directions will not only extend the practical applicability of the framework but also bridge the gap between simulation and hardware deployment, accelerating the realization of resilient, autonomous multi-robot systems.

\end{document}